\documentclass[preprint,12pt]{elsarticle}

\usepackage{graphicx}
\usepackage{amssymb}
\usepackage{lineno}
\usepackage{lipsum}
\makeatletter
\def\ps@pprintTitle{%
 \let\@oddhead\@empty
 \let\@evenhead\@empty
 \def\@oddfoot{}%
 \let\@evenfoot\@oddfoot}
\makeatother

\begin{document}

\begin{frontmatter}


\title{The Binary Vector Clock}

\author{Lum Ramabaja}
\ead{lum@bloomlab.io}

\address{Bloom Lab}

\begin{abstract}
The Binary Vector Clock is a simple, yet space-efficient algorithm for generating a partial order of transactions in account-based blockchain systems. The Binary Vector Clock solves the problem of order dependency in systems such as Ethereum, caused by the total order of transactions that come from the same address holder. The proposed algorithm has the same security as using regular transaction nonces, requires very little overhead, and can potentially result in a significant increase in throughput for systems like Ethereum. 
\end{abstract}


\end{frontmatter}

\section{Introduction}
There are generally two kinds of transaction models in blockchains: The UTXO model, and the account based model. The UTXO model was the first transaction model to be proposed and has many intriguing properties \cite{Nakamoto2008Bitcoin:System}. In this paper however, we are going to focus on the account based model, more exactly on the one implemented by Ethereum \cite{DR.GAVINWOODETHEREUM:LEDGER}. In the account based model, instead of having coins as unspent outputs like in the UTXO model, every participating node has an account, or a balance. When a transaction is created, the transaction's value is simply reduced from the owners account, and added to someone else's account. To understand the problem that the Binary Vector Clock tries to solve, let's first look at the structure of an Ethereum transaction and how the \textit{order} of transactions is determined.

\section{The order of Ethereum Transactions}
A transaction in Ethereum is essentially a message that gets signed by an account holder, also known as an externally owned account. Once a transaction gets created, it is broadcast to other nodes in the system, and eventually recorded by the Ethereum blockchain. The structure of an Ethereum transaction consist of: 
\begin{enumerate}
    \item A \textit{value} (the amount of ether we want to transfer).
    \item A \textit{recipient} (the address of the account to whom we want to send the transaction to).
    \item A \textit{gas price} (much like a transaction fee. The gas price shows how much of a fee the originator of the transaction is willing to pay).
    \item A \textit{gas limit} (the maximum fee that the originator is willing to pay).
    \item \textit{v,r,s} (the three ECDSA digital signature components to prove that the originator truly formed the transaction).
    \item A \textit{data} field (an optional field that can contain code, for when an account interacts with smart contracts).
    \item And the \textit{nonce} (an account specific counter. Whenever a transaction from the address holder gets confirmed, the counter increments).
\end{enumerate}{}

The nonce field is the field that is of particularly interest to us. The transaction nonce, not to be confused with the block nonce used for Proof of Work, is a scalar value that serves as a counter. The nonce shows the number of confirmed transactions that originated from the account. Having such a counter for each transaction has an interesting effect: It protects the user from transaction duplication. Let's see what would happen if transactions had no nonce, to better understand why having such a counter is so important: Let's say Alice sent Bob a completely valid transaction containing three ether. The signature turned out to be truly Alice's, and the transaction got recorded on the blockchain. Bob however turns out to have a bad moral compass and wants more money. Without a transaction nonce, there is nothing to stop Bob from "replaying" Alice's transaction, and claim again three ether. Bob could in fact repeat transmitting Alice's old transaction to the network, until he gets all of Alice's ether. Every time the transaction would be replayed, nodes in the system would think that it is a new transaction. In reality however, this is not what happens. By having a counter attached to the transaction, every transaction becomes unique. If let's say Alice's transaction has a nonce of 42, Bob will not be able to replay that transaction, as any new transaction coming from Alice would have to have a nonce greater than 42.

There is however also another important reason to have a nonce in an account-based transaction: We want to be able to determine the \textit{order} of transactions. Let's assume this time that Alice is sending two transactions, but the second transaction is dependent on the first one, i.e. running the second transaction before the first one is invalid (for whatever reason). In a centralized system this is no problem, one would simply confirm the first transaction first, and than continue with the second transaction. In a decentralized system however, nodes in the network might receive the second transaction before the first one. We cannot know in advance in which order nodes will perceive events. Without a counter, there would be no way for nodes in the network to tell which transaction comes first. If on the other hand the first transaction has a counter of 42 and the other transaction has the next counter (43), the order can be determined. If a node in the network thus receives the second transaction before the first one, it knows that it should ignore the second transaction, until the first transaction gets confirmed. 

This is a great feature, but it also has its shortcomings. If Alice were to send several transactions one after another, and one of the transactions does not get included in any block for some reason, e.g. the transaction turns out to be invalid, then none of the subsequent transactions get processed. Only after providing a transaction with the missing nonce, do all the other transactions get processed. 
This is no problem if every transaction depends on the previous one, but in most real-world applications that would not be the case. Many nodes have to create dozens of transactions in a short period of time, imposing an order dependency thus can result in transactions having to stay in mempools, even if they could have been processed sooner. The total order of transactions represents at the same time a great feature, and a serious scaling problem for account-based transaction models. In the following sections, I will present how we can overcome the problem of total order when processing transactions.

\section{Partial Orders and Join-Semilattices}
Before jumping straight to how the Binary Vector Clock works, it is necessary to have a good grasp of what a partial order is. All of us intuitively understand the idea of "total orders" - One is smaller than two, five is greater than four, etc. In order theory, a set is said to have a total order, if for any element $a$ and $b$, a comparison is possible, i.e. either $a \leq b$ or $a \geq b$. For example: Every transaction nonce for an address, is comparable to any other transaction nonce for that address. We thus can easily know which transaction happened-before another transaction, thanks to the total order of transactions. But what if it does not matter in which order some of our transactions get confirmed? If eight out of ten transactions generated from an address holder could in fact be confirmed in any desired order, it would be quite wasteful not to do so. This is however what happens in today's totally ordered account based transaction model.

It would thus be of enormous interest if we could somehow "capture" the transaction independence for address holders. This is where partial orders become useful. A partially ordered set, is a set in which only certain pairs of elements are comparable, i.e. one element precedes the other in the ordering, but not every pair of elements is necessarily comparable. 

\begin{figure}[ht]
\centering\includegraphics[width=0.4\linewidth]{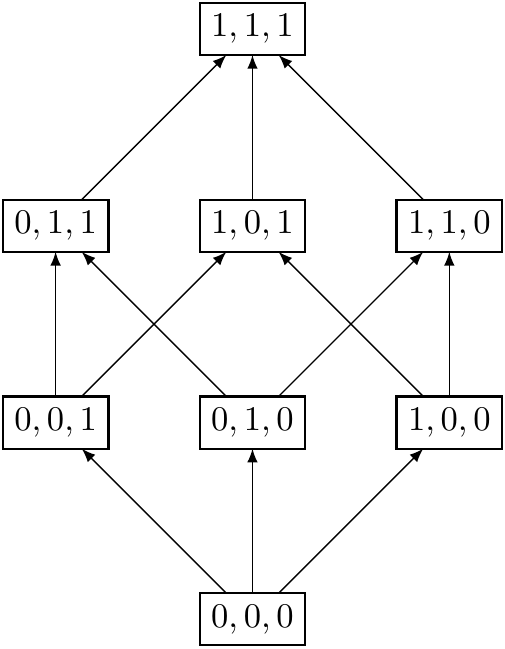}
\caption{A partially ordered diagram, also known as a join-semilattice.}
\label{fig:jointsemilattice}
\end{figure}

As an example to better understand what a partial order actually is, let's look at the join-semilattice in figure \ref{fig:jointsemilattice}. The diagram shows a set $S$ with eight vectors. We say that an element in $S$ ''happened-before'' another element, if and only if every value of vector $a$ is less than or equal to every corresponding value in vector $b$. For example: We can conclude that vector $(1,0,0)$ happened before vector $(1,1,0)$, because none of the values in vector $(1,0,0)$ are greater than in vector $(1,1,0)$ - We say that $(1,0,0)$ happened-before $(1,1,0)$. If on the other hand we try to compare vector $(1,1,0)$ and $(1,0,1)$, one can see that both vectors have values larger than the other vector at some indices. We say that this pair is \textit{not comparable}. One cannot determine which element occurred before the other one. Algorithms used in distributed systems, such as vector clocks, take advantage of partial orders. In the context of the distributed systems, having incomparable vectors, or "clocks", usually means that the events occurred concurrently, and thus have no information of one another. In the case of the Binary Vector Clock on the other hand, incomparableness between two transactions does not indicate concurrency, it indicates that they occur \textit{independently}.

\section{The Binary Vector Clock}
Let's imagine that instead of a nonce (i.e. counter) for a transaction, we have a counter \textit{and} a very small bit array (for the sake of a better explanation, let's stick to three bits, like the vectors in figure \ref{fig:jointsemilattice}). Alice's Binary Vector Clock is initially set to $(0, [0,0,0])$ (where the first element represents the counter and the second element the bit array). For simplicity, I will refer to the Binary Vector Clock from now on as a "timestamp". Now let's say Alice wants to send three transactions one after another. Alice however knows that her second transaction is dependent on her first transaction, but her third transaction has no logical dependency to the two first transactions. Having this information, Alice can do something clever: Instead of incrementing her counter for each transaction, she increments one of the bits in her bit array. Let's say the first transaction has the timestamp $(0,[0,0,1])$, the second transaction has the timestamp  $(0,[0,1,1])$, and the third timestamp is $(0,[1,0,0])$. All three transactions were send one after another to the network. Any validator receiving the transactions can independently know in what order the transactions need to be confirmed (or if any order exists at all). Validators first look at the counter, the counter tells a validator if the transaction is in the right "epoch" (more on that in a bit). If the counter is equal to the previously confirmed transaction from that address, the bit array is checked. As the bit array of the first and third transaction are not comparable (no order can be determined), even if the first transaction turns out to be invalid for some reason, the third transaction can still be processed by the validators. This is because both timestamps are indicating "independentness", there is no "happened-before" relationship between them. The second and the first transaction on the other hand do have a "happened-before" relationship. When looking at the bit array of the second transaction, we can conclude that it must have happened after the first transaction. If a validator thus would receive the third transaction and the second transaction, but not the first transaction for some reason, it would know that the third transaction can be processed, but the second transaction not, as it depends on a prior transaction (the first transaction). If a transaction gets confirmed, the address' Binary Vector Clock gets simply added with the newly confirmed timestamp. Taking again the three transactions from the previous scenario as an example, if Alice's initial timestamp was $(0,[0,0,0])$, and her first and third transactions get confirmed, her new timestamp would be $(0,[1,0,1])$. Once all the bits in the bit array are turned to one, we can increment the timestamp's counter, and set the bit array to zero again. We call this shift an "epoch". 

Up to this point, some of the readers might have already thought something in the lines of: But what if Alice has only one ether, and she creates three independent transactions, each spending one ether? It is important to remember that this is an issue only if transactions would be processed concurrently, which is not the case with the Binary Vector Clock technique. In cases like the one mentioned above, transactions would be treated the same way today's transactions get treated, if they were to have the same nonce. Today, with the nonce approach, if transactions have the same nonce, one of the transactions would get confirmed (depending on the block creator) and the rest of the transactions would become invalid. In the case of the Binary Clock, one of Alice's transactions (depending on the block creator) would get confirmed, while the rest of the transactions would simply be considered invalid, regardless of their order independency. 

\section{The inevitable total order during epoch jumps}
It is important to note that there is nonetheless an inevitable transaction processing dependency when shifting from one epoch to the next. Transactions from one epoch can only be processed independently, after the transactions of the previous epoch were already processed. In other words if Alice were to send three other transactions one after the other, where the first transaction would have a timestamp of $(0,[1,1,1])$, second transaction $(1,[1,0,0])$, and third transaction $(1,[0,0,1])$, even if all three transactions are completely independent from one another, the second and third transactions will not be able to get processed without the first one being confirmed first. This is because these transactions occurred during an epoch "jump", i.e. the Binary Vector Clock gets incremented, and the bit array becomes set to zero. The transactions in the new epoch cannot know if they are comparable or not with the transaction from the previous epoch, forcing a momentary total order. I argue however that the space-wise inexpensive nature of the Binary Vector Clock, and its property to handle partial orders, makes it an attractive technique for the account-based transaction model, even in the case of momentary order dependencies between epoch jumps.

\section{Conclusion}
In This paper I introduced the Binary Vector Clock, a memory-wise inexpensive partially ordered counter for account-based transactions, that solves the issue of order dependency when processing transactions. Note that the Binary Vector Clock does not suggest the concurrent processing of transactions in Ethereum. Doing so would in fact introduce many possible attack vectors to the system. It only specifies which transactions can be processed independently, and which ones depend on a prior transaction confirmation. If for example an address generates $N$ transactions one after another, and the first transaction fails, the subsequent transactions are still able to get processed and confirmed by the blockchain. This is not the case in today's approach with transaction nonces. In today's approach, if the first transaction fails for some reason, all of the other transactions would need to be ignored until the gap in the nonce becomes filled. The Binary Vector Clock overcomes the issue by introducing a partial order between transactions of the same address holder. Using the Binary Vector Clock as a substitution for the transaction nonce gives more freedom to the user in determining transaction orders. The Binary Vector Clock allows the user to specify if a transaction can be processed  independently from other transactions, or if it should be queued until a certain transaction gets confirmed. I argue that this ability has important implications for blockchain systems. Considering that transactions in blockchain systems most likely follow a pareto distribution (the majority of transactions are generated by very few nodes), introducing an inexpensive technique that allows for independent processing of transactions, could potentially increase the scaling capability of Ethereum and other account-based blockchains significantly.

\bibliographystyle{model1-num-names}

\begin{thebibliography}{2}
    \expandafter\ifx\csname natexlab\endcsname\relax\def\natexlab#1{#1}\fi
    \providecommand{\bibinfo}[2]{#2}
    \ifx\xfnm\relax \def\xfnm[#1]{\unskip,\space#1}\fi
    \bibitem[{Nakamoto(2008)}]{Nakamoto2008Bitcoin:System}
    \bibinfo{author}{S.~Nakamoto}, \bibinfo{title}{{Bitcoin: A Peer-to-Peer
      Electronic Cash System}}, \bibinfo{type}{Technical Report},
      \bibinfo{year}{2008}.
    \bibitem[{{DR. GAVIN WOOD}(????)}]{DR.GAVINWOODETHEREUM:LEDGER}
    \bibinfo{author}{{DR. GAVIN WOOD}}, \bibinfo{title}{{ETHEREUM: A SECURE
      DECENTRALISED GENERALISED TRANSACTION LEDGER}}, \bibinfo{type}{Technical
      Report}, ????
    
    \end{thebibliography}

\end{document}